\documentclass[twocolumn,preprintnumbers,amsmath,amssymb]{revtex4}
\usepackage{dcolumn}
\usepackage{bm}
\usepackage{graphicx}
\usepackage{amsmath}
\usepackage{afterpage}
\usepackage{epstopdf}
\begin{document}

\title{Stabilizing correlated pair tunneling of spin-orbit-coupled bosons in a non-Hermitian driven double well}
\author{\ Miaoqian Lu$^{1}$, \ Xinzhou Guan$^{1}$, \ Mohan Xia$^{1}$, \ Wenjuan Li$^{2}$, \ Jincheng Hu$^{1}$, \ Xinyue Zhang$^{1}$, and Yunrong Luo$^{1}$\footnote{Corresponding author: lyr\underline{ }1982@hunnu.edu.cn}}
\affiliation{$^{1}$Key Laboratory of Low-dimensional Quantum Structures and Quantum Control of Ministry
of Education, Key Laboratory for Matter Microstructure and Function of Hunan Province, and Hunan Research Center of the Basic Discipline for Quantum Effects and Quantum Technologies, School of Physics and Electronics, Hunan Normal University, Changsha 410081, China\\
$^{2}$School of Physics and Electronic Information Engineering, Ningxia Normal University, Guyuan, Ningxia 756000, China}
\begin{abstract}
We present an analytical framework for stabilizing second-order correlated tunneling of two spin-orbit-coupled bosons in a periodically driven non-Hermitian double-well potential. By combining Floquet theory with multiple-scale asymptotic analysis, we derive effective second-order dynamics and exact quasienergy spectra in the strongly interacting regime. Our analysis reveals distinct stability mechanisms of correlated pair tunneling for three fundamental tunneling channels: interwell spin-conserving, interwell spin-flipping, and intrawell spin-flipping. For balanced gain and loss, we identify discrete, well-defined parameter regions where stable pair tunneling emerges, with the spin-flipping channel exhibiting a characteristic symmetry absent in its spin-conserving counterpart. Under unbalanced gain-loss conditions, stability is achieved only when the gain and loss coefficients satisfy specific parametric relations, enabling dissipation-controlled tunneling. Most notably, stable intrawell spin-flipping, while inherently unstable for an initial Fock state, becomes accessible when the system is prepared in a coherent superposition state, thereby revealing that initial-state coherence can serve as a control parameter for dynamical stability in non-Hermitian systems. These results expand the possibilities for controlling correlated tunneling in many-body systems with engineered dissipation.
\end{abstract}

\maketitle

\section{Introduction}

Periodically driven double-well potentials have long served as a paradigmatic testbed for exploring coherent control of tunneling, interaction effects, and few-body correlations in ultracold atomic gases\cite{Winkler441, Tai546, Strohmaier104, Hai82, Zhou15, Longhi86}. Within this framework, a rich variety of phenomena have been uncovered, including coherent destruction of tunneling\cite{Grossmann67, Grossmann84}, the formation of repulsively bound pairs (doublons)\cite{Winkler441}, and Floquet-engineered Hamiltonians\cite{Zhou15}. More recently, the experimental realization of synthetic spin-orbit coupling (SOC) for ultracold atoms using Raman laser schemes has opened unprecedented avenues for investigating spin-resolved quantum dynamics, topological states of matter, and synthetic gauge fields\cite{Bernevig2006, zutic2004, Kato2004, Dalibard2011, Lin2009, Lin2011}. The combination of SOC with double-well potentials has revealed intriguing phenomena, such as spin Josephson effects\cite{Garcia89, Zhang85} and SOC-mediated localization in bosonic junctions\cite{Wu26}.

Concurrently, the study of non-Hermitian quantum systems has fundamentally reshaped our understanding of dynamics in open environments and systems with engineered gain and loss. A pivotal development in this field is the concept of parity-time (PT) symmetry, which demonstrates that real spectra and consequently stable dynamics can be preserved even in the presence of balanced gain and loss\cite{Bender1998, Bender1999, Ganainy11}. This insight has led to the discovery of numerous exotic phenomena, including exceptional points\cite{Miri2019, Zhang2024}, the non-Hermitian skin effect\cite{Yao2018, gong124, Wang2024}, and PT-symmetric lasing\cite{Ganainy11}, which have been experimentally realized across various platforms including photonic, acoustic, and cold-atom systems\cite{gao50, wang14, Ding2016, li10, liang129}.

The convergence of three research frontiers, namely, periodic driving, SOC, and non-Hermitian physics, within well-controlled few-body systems presents a natural and promising avenue for discovering novel quantum phenomena that transcend the bounds of conventional Hermitian physics. However, despite significant progress in each individual direction, the vast majority of prior work has been confined to either purely Hermitian systems\cite{luo93, luo56, Li2022, Yu90} or single-particle dynamics in non-Hermitian settings\cite{tang55, Luo2020, Xie2023}. The critical question of how correlated pair tunneling with interparticle interaction, a fundamentally many-body process, can be stabilized in a non-Hermitian system with SOC remains largely unexplored. This gap is not merely a technical nuance; it addresses a fundamental question about the survivability of many-body correlations in the presence of dissipation and gain, which are unavoidable in any realistic quantum simulation platform.

In this work, we address this fundamental question by systematically investigating the stability of second-order tunneling processes for two spin-orbit-coupled bosons confined in a periodically driven non-Hermitian double well. By developing an analytical framework that captures the interplay between SOC, driving, and non-Hermiticity, we aim to uncover not just whether, but how correlated pair tunneling can be preserved and controlled in an open environment. Our analysis focuses on three distinct dynamical pathways: interwell spin-conserving tunneling, interwell spin-flipping tunneling, and intrawell spin-flipping. By employing a powerful combination of Floquet theory and multiple-scale asymptotic analysis, we derive effective second-order equations of motion and obtain analytical expressions for the Floquet quasienergy spectra. This approach allows us to precisely identify stable parameter regimes under both balanced and unbalanced gain-loss conditions. Our results reveal that stable pair tunneling can be achieved in well-defined, discrete parameter regions, with the spin-flipping channel exhibiting a unique symmetry property not shared by its spin-conserving counterpart. Furthermore, we find that stable intrawell spin-flipping is contingent upon preparing the system in a coherent superposition state, highlighting the critical role of quantum coherence in dissipative spin dynamics. These findings provide a theoretical framework for controlling correlated quantum tunneling in non-Hermitian quantum systems.

\section{The model and multiple-scale asymptotic analysis}
We consider two ultracold bosons with synthetic spin-orbit coupling confined to a driven non-Hermitian double-well potential. The system is described by an extended two-site Bose-Hubbard Hamiltonian\cite{Yu90, tang55, Luo2020, Xie2023, tao136}
\begin{align}
\hat{H}&=-\nu(\hat{a}^{\dagger}_{1}e^{-i\pi\alpha\hat{\sigma}_{y}}\hat{a}_{2}+H.c.)+\frac{\delta}{2}\sum_{j=1,2}(\hat{a}^{\dagger}_{j\uparrow}\hat{a}_{j\downarrow}+H.c.)\nonumber\\
&+\sum_{j=1,2}[\Omega (t)\hat{a}^{\dagger}_{j}\hat{\sigma}_{z}\hat{a}_{j}+f_{j}(t)\hat{a}^{\dagger}_{j}\hat{a}_{j}]\nonumber\\
&+\sum_{j=1,2}\sum_{\sigma\sigma^{\prime}}U_{\sigma\sigma^{\prime}}\hat{a}^{\dagger}_{j\sigma}\hat{a}^{\dagger}_{j\sigma^{\prime}}\hat{a}_{j\sigma^{\prime}}\hat{a}_{j\sigma}.
\end{align}
Here $\hat{a}_{j}^{\dagger}=(\hat{a}_{j,\uparrow}^{\dagger},\hat{a}_{j,\downarrow}^{\dagger})$ and $\hat{a}_{j}=(\hat{a}_{j,\uparrow},\hat{a}_{j,\downarrow})^{\mathrm{T}}$ (the superscript T stands for the transpose) are the creation and annihilation operators for a boson with pseudospin $\sigma$ ($\sigma$=$\uparrow,\downarrow$) in well $j$ ($j$=$1,2$). The parameter $\nu$ denotes the spin-independent tunneling rate, $\alpha$ characterizes the strength of the SOC, and $\hat{\sigma}_{y}$ is the Pauli matrix. The strength of the Raman coupling, which drives transitions between the spin states, is given by $\delta$. \emph{H.c.} denotes the Hermitian conjugate of the preceding term. The time-dependent Zeeman field is $\Omega(t)=\Omega cos(\omega t)$ with driving frequency $\omega$ and amplitude $\Omega$. The terms $f_{1}(t)=f cos(\omega t)+i\beta_{1}$ and $f_{2}(t)=-f cos(\omega t)-i\beta_{2}$ describe an ac driving force with amplitude $f$ and a non-Hermitian gain-loss profile $\beta_j$\cite{tao136}. For $\beta_j >0$, well 1 is a source of particles (gain) and well 2 is a sink (loss). $U_{\sigma,\sigma^\prime}$ represents the on-site interaction strength between two atoms, which can be adjusted by Feshbach resonances in the experiment\cite{Chin82}. For simplicity and without loss of generality, we set $U_{\uparrow,\uparrow}=U_{\downarrow,\downarrow}=U_{1}$ and $U_{\uparrow,\downarrow}=U_{2}$. Throughout this paper, we work in units where $\hbar=1$. All parameters ($\nu,\delta,\Omega,\omega,f$) are expressed in units of a reference frequency $\omega_{0}=0.1E_{R}$, with $E_{R}=k_{R}^{2}/2M=22.5$kHz being the recoil energy\cite{Lin2011, Yu90}. Time $t$ is measured in units of $\omega_{0}^{-1}$.

The Hilbert space for two particles is spanned by the Fock states $|n_{1,\uparrow} n_{1,\downarrow} n_{2,\uparrow} n_{2,\downarrow}\rangle$, where $n_{j,\sigma}$ represents the number of spin-$\sigma$ particles in well $j$. The general quantum state can be expanded as
\begin{align}
|\psi(t)\rangle&=c_{1}(t)|1100\rangle+c_{2}(t)|2000\rangle+c_{3}(t)|0200\rangle \nonumber\\
&+c_{4}(t)|0011\rangle+c_{5}(t)|0020\rangle+c_{6}(t)|0002\rangle\nonumber\\
&+c_{7}(t)|1010\rangle+c_{8}(t)|1001\rangle+c_{9}(t)|0110\rangle\nonumber\\
&+c_{10}(t)|0101\rangle.
\end{align}
Here $c_{k}(t) (k=1,2,3,...,10)$ are the time-dependent probability amplitudes for the corresponding Fock states, and $P_{k}(t)=|c_{k}(t)|^{2}$ are the associated occupation probabilities.

In this work, we focus on three specific tunneling channels: interwell spin-conserving, interwell spin-flipping, and intrawell spin-flipping. These channels are selected because they represent the fundamental building blocks of correlated pair dynamics in the presence of SOC and driving. By tuning the SOC strength $\alpha$ and the system parameters, the system can be configured to predominantly activate one of these channels while suppressing others. For example, setting $\alpha=1$ ($\alpha=0.5$) and $\delta=0$ isolates interwell spin-conserving (spin-flipping) tunneling\cite{Yu90, Luo2020}. Each channel exhibits distinct stability behaviors under non-Hermitian conditions, providing a comprehensive picture of how dissipation and gain affect correlated tunneling.

In the following, we focus on the tunneling dynamics of bound states (doublons) in the limit of strong on-site interactions and high-frequency driving. To this end, we perform a multiple-scale asymptotic analysis of the Bose-Hubbard model in the strong interaction regime $(U_{j}\gg\nu)$ and under high-frequency driving $(\omega\gg\nu)$\cite{Longhi86}. To ensure the validity of our perturbative approach, we also require that the condition $2U_{j}\neq m\omega$ ($m=0,1,2\dots$) is met\cite{Zhou15, Wu26}, thus avoiding resonant processes that would invalidate a simple second-order expansion. We introduce the small parameters $\epsilon=\frac{\nu}{\omega}$ and $\theta=\frac{\delta}{\omega}$, and define the normalized time variable $\tau=\omega t$. For each of the three tunneling channels of interest, we will choose an appropriate initial state and derive the effective second-order dynamics and the corresponding quasienergies.

\subsection{Interwell spin-conserving tunneling}
For integer $\alpha$ (e.g., $\alpha=1$) and $\delta=0$, the Hamiltonian (1) reduces to
\begin{equation}
\begin{aligned}
\hat{H}&=\nu(\hat{a}^{\dagger}_{1\uparrow}\hat{a}_{2\uparrow}+\hat{a}^{\dagger}_{1\downarrow}\hat{a}_{2\downarrow}+H.c.)\\
&+\sum_{j=1,2}[\Omega (t)\hat{a}^{\dagger}_{j}\hat{\sigma}_{z}\hat{a}_{j}+f_{j}(t)\hat{a}^{\dagger}_{j}\hat{a}_{j}]\\
&+\sum_{j=1,2}\sum_{\sigma\sigma^{\prime}}U_{\sigma\sigma^{\prime}}\hat{a}^{\dagger}_{j\sigma}\hat{a}^{\dagger}_{j\sigma^{\prime}}\hat{a}_{j\sigma^{\prime}}\hat{a}_{j\sigma}.
\end{aligned}
\end{equation}
Such that the system only exhibits spin-conserving tunneling. When the system is initially prepared in the state $|0020\rangle$ (two spin-up atoms in the right well) , the dynamics is confined to the subspace $\{|0020\rangle,|1010\rangle,|2000\rangle\}$\cite{Wu26}. The wavefunction ansatz is
\begin{align}
|\psi(t)\rangle=c_{5}|0020\rangle+c_{7}(t)|1010\rangle+c_{2}(t)|2000\rangle.
\end{align}
Substituting equations (3) and (4) into Schr\"{o}dinger equation $i\frac{\partial|\psi(t)\rangle}{\partial t}=\hat{H}(t)|\psi(t)\rangle$, yields a set of coupled equations for the amplitudes
\begin{equation}
\begin{aligned}
i\frac{dc_{5}}{dt}&=(2\Omega-2f)\cos(\omega t)c_{5}(t)+2U_{1}c_{5}(t)\\
&+\sqrt{2}\nu c_{7}(t)-2i\beta_{2}c_{5}(t),\\
i\frac{dc_{7}}{dt}&=\sqrt{2}\nu c_{2}(t)+\sqrt{2}\nu c_{5}(t)\\
&+2\Omega\cos(\omega t)c_{7}(t)+(i\beta_{1}-i\beta_{2})c_{7}(t),\\
i\frac{dc_{2}}{dt}&=(2\Omega+2f)\cos(\omega t)c_{2}(t)+2U_{1}c_{2}(t)\\
&+\sqrt{2}\nu c_{7}(t)+2i\beta_{1}c_{2}(t).
\end{aligned}
\end{equation}
Making the transformation, $c_{5}(t)=C_{5}(t)e^{-i[\frac{2\Omega-2f}{\omega}sin(\omega t)+2U_{1}t]}$, $c_{7}(t)=C_{7}(t)e^{-i[\frac{2\Omega}{\omega}sin(\omega t)]}$, $c_{2}(t)=C_{2}(t)e^{-i[\frac{2\Omega+2f}{\omega}sin(\omega t)+2U_{1}t]}$, recalling that $\epsilon=\frac{\nu}{\omega}$ and $\tau=\omega t$, then we rewrite equation (5) as
\begin{equation}
\begin{aligned}
i\frac{dC_{5}}{d\tau}&=\sqrt{2}\epsilon e^{-i\frac{2f}{\omega}sin\tau+i\frac{2U_{1}}{\omega}\tau}C_{7}-2i\frac{\beta_{2}}{\omega}C_{5},\\
i\frac{dC_{7}}{d\tau}&=\sqrt{2}\epsilon e^{-i\frac{2f}{\omega}sin\tau-i\frac{2U_{1}}{\omega}\tau}C_{2}\\
&+\sqrt{2}\epsilon e^{i\frac{2f}{\omega}sin\tau-i\frac{2U_{1}}{\omega}\tau}C_{5}\\
&+(i\frac{\beta_{1}}{\omega}-i\frac{\beta_{2}}{\omega})C_{7},\\
i\frac{dC_{2}}{d\tau}&=\sqrt{2}\epsilon e^{i\frac{2f}{\omega}sin\tau+i\frac{2U_{1}}{\omega}\tau}C_{7}+2i\frac{\beta_{1}}{\omega}C_{2}.
\end{aligned}
\end{equation}
We expand $C_{k}(k=2,5,7)$ as a power-series of $\epsilon$
\begin{align}
C_{k}(\tau)=C_{k}^{(0)}(\tau)+\epsilon C_{k}^{(1)}(\tau)+\epsilon^{2}C_{k}^{(2)}(\tau)+\dots
\end{align}
At the same time, we introduce multiple time scales $\tau_{0}=\tau$, $\tau_{1}=\epsilon\tau$, $\tau_{2}=\epsilon^{2}\tau$, ..., and then replace the time derivatives by the expansion
\begin{align}
\frac{d}{d\tau}=\partial_{\tau_{0}}+\epsilon\partial_{\tau_{1}}+\epsilon^{2}\partial_{\tau_{2}}+\dots
\end{align}
Substituting equations (7) and (8) into equation (6), and collecting the terms of the same order, we obtain a hierarchy of approximation equations of different orders in $\epsilon$. At the order $\epsilon^{0}$, we find
\begin{equation}
\begin{aligned}
i\frac{\partial C_{5}^{(0)}}{\partial \tau_{0}}=0, C_{5}^{(0)}=A_{5}(\tau_{1},\tau_{2},\ldots),\\
i\frac{\partial C_{7}^{(0)}}{\partial \tau_{0}}=0, C_{7}^{(0)}=A_{7}(\tau_{1},\tau_{2},\ldots),\\
i\frac{\partial C_{2}^{(0)}}{\partial \tau_{0}}=0, C_{2}^{(0)}=A_{2}(\tau_{1},\tau_{2},\ldots),
\end{aligned}
\end{equation}
where the amplitudes $A_{k}(\tau_{1},\tau_{2},\ldots)$, $k=2,5,7$ are functions of the slow time variables $\tau_{1},\tau_{2},\ldots$, but independent of the fast time variables $\tau_{0}$. At order $\epsilon^{1}$ we have
\begin{equation}
\begin{aligned}
i\frac{\partial C_{5}^{(1)}}{\partial \tau_{0}}&=-i\partial_{\tau_{1}}A_{5}+\sqrt{2}A_{7}e^{-i\frac{2f}{\omega}\sin\tau+i\frac{2U_{1}}{\omega}\tau}\\
&-2i\frac{\beta_{2}}{\nu}A_{5},\\
i\frac{\partial C_{7}^{(1)}}{\partial \tau_{0}}&=-i\partial_{\tau_{1}}A_{7}+\sqrt{2}A_{2}e^{-i\frac{2f}{\omega}\sin\tau-i\frac{2U_{1}}{\omega}\tau}\\
&+\sqrt{2}A_{5}e^{i\frac{2f}{\omega}\sin\tau-i\frac{2U_{1}}{\omega}\tau}+i(\frac{\beta_{1}}{\nu}-\frac{\beta_{2}}{\nu})A_{7},\\
i\frac{\partial C_{2}^{(1)}}{\partial \tau_{0}}&=-i\partial_{\tau_{1}}A_{2}+\sqrt{2}A_{7}e^{i\frac{2f}{\omega}\sin\tau+i\frac{2U_{1}}{\omega}\tau}\\
&+2i\frac{\beta_{1}}{\nu}A_{2}.
\end{aligned}
\end{equation}
For the convenience of our discussion, we simplify equation (10) as $i\partial C_{k}^{(1)}/\partial\tau_{0}=-i\partial_{\tau_{1}}A_{k}+G_{k}^{(1)}(\tau_{0})$ for $k=2,5,7$. To avoid the occurrence of secular growing terms in the solution $C_{k}^{(1)}$, the solvability conditions must be satisfied
\begin{equation}
\begin{aligned}
i\partial_{\tau_{1}}A_{5}=\overline{G_{5}^{(1)}(\tau_{0})},\\
i\partial_{\tau_{1}}A_{7}=\overline{G_{7}^{(1)}(\tau_{0})},\\
i\partial_{\tau_{1}}A_{2}=\overline{G_{2}^{(1)}(\tau_{0})}.
\end{aligned}
\end{equation}
Throughout our paper, the overline denotes the time average with respect to the fast time variable $\tau_{0}$. The solvability at order $\epsilon^{1}$ then gives
\begin{equation}
\begin{aligned}
i\partial_{\tau_{1}}A_{5}&=-2i\frac{\beta_{2}}{\nu}A_{5},\\
i\partial_{\tau_{1}}A_{7}&=i(\frac{\beta_{1}}{\nu}-\frac{\beta_{2}}{\nu})A_{7},\\
i\partial_{\tau_{1}}A_{2}&=2i\frac{\beta_{1}}{\nu}A_{2}.
\end{aligned}
\end{equation}
According to $C_{k}^{(1)}=-i\int[G_{k}^{(1)}(\tau_{0})-\overline{G_{k}^{(1)}(\tau_{0})}]d\tau_{0}$, the amplitudes $C_{5}^{(1)}$, $C_{7}^{(1)}$, $C_{2}^{(1)}$ at order $\epsilon^{1}$ are given by
\begin{equation}
\begin{aligned}
C_{5}^{(1)}&=-i\sqrt{2}A_{7}F_{1}(\tau_{0}),\\
C_{7}^{(1)}&=-i\sqrt{2}A_{2}F_{2}^*(\tau_{0})-i\sqrt{2}A_{5}F_{1}^*(\tau_{0}),\\
C_{2}^{(1)}&=-i\sqrt{2}A_{7}F_{2}(\tau_{0}),
\end{aligned}
\end{equation}
with
\begin{equation}
\begin{aligned}
F_{1}(\tau_{0})&=\sum_{r}\frac{J_{r}(\frac{2f}{\omega})e^{i(-r+\frac{2U_{1}}{\omega})\tau_{0}}}{i(-r+\frac{2U_{1}}{\omega})},\\
F_{2}(\tau_{0})&=\sum_{r}\frac{J_{r}(\frac{2f}{\omega})e^{i(r+\frac{2U_{1}}{\omega})\tau_{0}}}{i(r+\frac{2U_{1}}{\omega})},
\end{aligned}
\end{equation}
and $J_{r}$ being the $r$-th order Bessel function of the first kind.
At the next order $\epsilon^{2}$, we have
\begin{equation}
\begin{aligned}
i\frac{\partial C_{5}^{(2)}}{\partial \tau_{0}}&=-i\partial_{\tau_{2}}A_{5}-i\partial_{\tau_{1}}C_{5}^{(1)}+G_{5}^{(2)}(\tau_{0}),\\
i\frac{\partial C_{7}^{(2)}}{\partial \tau_{0}}&=-i\partial_{\tau_{2}}A_{7}-i\partial_{\tau_{1}}C_{7}^{(1)}+G_{7}^{(2)}(\tau_{0}),\\
i\frac{\partial C_{2}^{(2)}}{\partial \tau_{0}}&=-i\partial_{\tau_{2}}A_{2}-i\partial_{\tau_{1}}C_{2}^{(1)}+G_{2}^{(2)}(\tau_{0}),
\end{aligned}
\end{equation}
with
\begin{equation}
\begin{aligned}
G_{5}^{(2)}(\tau_{0})&=\sqrt{2}C_{7}^{(1)}e^{-i\frac{2f}{\omega}\sin\tau+i\frac{2U_{1}}{\omega}\tau},\\
G_{7}^{(2)}(\tau_{0})&=\sqrt{2}C_{2}^{(1)}e^{-i\frac{2f}{\omega}\sin\tau-i\frac{2U_{1}}{\omega}\tau}\\
&+\sqrt{2}C_{5}^{(1)}e^{i\frac{2f}{\omega}\sin\tau-i\frac{2U_{1}}{\omega}\tau},\\
G_{2}^{(2)}(\tau_{0})&=\sqrt{2}C_{7}^{(1)}e^{i\frac{2f}{\omega}\sin\tau+i\frac{2U_{1}}{\omega}\tau}.
\end{aligned}
\end{equation}
In order to avoid the occurrence of secularly growing terms in the solutions $C_{k}^{(2)}$, the following solvability conditions must be satisfied
\begin{equation}
\begin{aligned}
i\partial_{\tau_{2}}A_{5}&=\overline{G_{5}^{(2)}(\tau_{0})}=2(A_{2}\rho_{2}+A_{5}\rho_{1}),\\
i\partial_{\tau_{2}}A_{7}&=\overline{G_{7}^{(2)}(\tau_{0})}=-4A_{7}\rho_{1},\\
i\partial_{\tau_{2}}A_{2}&=\overline{G_{2}^{(2)}(\tau_{0})}=2(A_{2}\rho_{1}+A_{5}\rho_{2}).
\end{aligned}
\end{equation}
The effective coupling parameters are given by
\begin{equation}
\begin{aligned}
\rho_{1}&=\sum_{p}\frac{J_{p}^{2}(\frac{2f}{\omega})}{p+\frac{2U_{1}}{\omega}},\\
\rho_{2}&=\sum_{p}\frac{J_{p}(\frac{2f}{\omega})J_{-p}(\frac{2f}{\omega})}{p+\frac{2U_{1}}{\omega}},
\end{aligned}
\end{equation}
with $p+\frac{2U_{1}}{\omega}\neq0$. Thus the evolution of the amplitudes $A_{k}$ up to the second-order long time scale is given by
\begin{align}
i\frac{dA_{k}}{d\tau}=i\frac{\partial A_{k}}{\partial \tau_{0}}+i\epsilon\frac{\partial A_{k}}{\partial \tau_{1}}+i\epsilon^{2}\frac{\partial A_{k}}{\partial \tau_{2}},(k=2,5,7).
\end{align}
Substituting equations (9), (12), and (17) into equation (19), and returning to the original time variable $t$, we have
\begin{equation}
\begin{aligned}
i\frac{dA_{5}}{dt}&=\frac{2\nu^{2}}{\omega}(A_{2}\rho_{2}+A_{5}\rho_{1})-2i\beta_{2}A_{5},\\
i\frac{dA_{7}}{dt}&=-\frac{4\nu^{2}}{\omega}A_{7}\rho_{1}+(i\beta_{1}-i\beta_{2})A_{7},\\
i\frac{dA_{2}}{dt}&=\frac{2\nu^{2}}{\omega}(A_{2}\rho_{1}+A_{5}\rho_{2})+2i\beta_{1}A_{2}.
\end{aligned}
\end{equation}

Equation (20) reveals a key feature: the dynamics of the two doublon states ($A_{2}(t)$ and $A_{5}(t)$) are decoupled from the unpaired state ($A_{7}(t)$). This confirms that the second-order process indeed describes the correlated tunneling of bound pairs. The validity of this description hinges on
$2U_{1}/\omega$ being sufficiently far from any integer, ensuring that $\rho_{1}$ and $\rho_{2}$  are well-behaved and the perturbation theory is accurate.

According to the Floquet theroy, the solutions of the periodic time-dependent Schr\"{o}dinger equation can be written as $|\psi_{k}(t)\rangle=e^{-iE_{k}t}|\varphi_{k}(t)\rangle$, with $|\varphi_{k}(t)\rangle$ being the Floquet states and $E_{k}$ Floquet quasienergies. We can construct the Floquet states by setting $A_{5}(t)=B_{5}e^{-i(E-2U_{1})t}$, $A_{7}(t)=B_{7}e^{-iEt}$ and $A_{2}(t)=B_{2}e^{-i(E-2U_{1})t}$, where $B_{k}$ are constant. Substituting into equation (20) yields a time-independent eigenvalue problem, whose solution provides the three Floquet quasienergies:
\begin{equation}
\begin{aligned}
E_{1}&=i(\beta_{1}-\beta_{2})-\frac{4\nu^{2}\rho_{1}}{\omega},\\
E_{2}&=2U_{1}+i(\beta_{1}-\beta_{2})+\frac{2\nu^{2}\rho_{1}}{\omega}-\zeta,\\
E_{3}&=2U_{1}+i(\beta_{1}-\beta_{2})+\frac{2\nu^{2}\rho_{1}}{\omega}+\zeta,
\end{aligned}
\end{equation}
where we have defined
\begin{align}
\zeta=\sqrt{(\frac{2\nu^{2}\rho_{2}}{\omega})^{2}-(\beta_{1}+\beta_{2})^{2}}.
\end{align}

Before proceeding to the stability analysis, it is instructive to examine the structure of the quasienergies in equation (21). The term $2U_{1}$ represents the bare interaction energy of the doublon. The term $i(\beta_{1}-\beta_{2})$  is the net gain/loss experienced by the system. The contributions from
$\rho_{1}$ and $\rho_{2}$ are the Floquet-renormalized second-order tunneling amplitudes. Crucially, $\zeta$ contains the competition between the coherent pair tunneling amplitude ($\propto \frac{\nu^{2}\rho_{2}}{\omega}$) and the total dissipation rate ($\beta_{1}+\beta_{2}$). It is this competition, encapsulated in the radicand of $\zeta$, that ultimately dictates the system's stability.

In a non-Hermitian system, stability is determined by the imaginary parts of the quasienergies, Im($E_{k}$). The system is stable under two distinct conditions\cite{Luo2020}:

Condition (i). Balanced gain-loss ($\beta_{1}=\beta_{2}$): Stability requires all Im($E_{k}$)=0, meaning all quasienergies are real\cite{Luo95}. This implies Im($\zeta$)=0.

Condition (ii). Unbalanced gain-loss ($\beta_{1}<\beta_{2}$): Stability is achieved if some Im($E_{k}$)=0 and the others are negative\cite{Xiao85, Zhou384}. This generally requires a specific relation between parameters.

\subsubsection{Stability under balanced gain and loss}

For balanced gain and loss, namely, $\beta_{1}=\beta_{2}=\beta$, the quasienergies in equation (21) simplify to
\begin{equation}
\begin{aligned}
E_{1}&=-\frac{4\nu^{2}\rho_{1}}{\omega},\\
E_{2}&=2U_{1}+\frac{2\nu^{2}\rho_{1}}{\omega}-\zeta,\\
E_{3}&=2U_{1}+\frac{2\nu^{2}\rho_{1}}{\omega}+\zeta,
\end{aligned}
\end{equation}
with $\zeta=\sqrt{(\frac{2\nu^{2}\rho_{2}}{\omega})^{2}-(2\beta)^{2}}$.

Based on the condition (i), the imaginary parts of the quasienergies vanish only when Im($\zeta$)=0, i.e., when the coherent coupling dominates over dissipation: $\mid\frac{2\nu^{2}\rho_{2}}{\omega}\mid\geq 2\beta$.
In figure 1(a), we set the parameters $\nu=\alpha=1$, $\delta=0$, $\omega=\Omega=40$, and $\beta_{1}=\beta_{2}=\beta=0.01$ to plot Im($\zeta$) as a function of $2U_{1}/\omega$ and $2f/\omega$, where the red line represents the boundary between Im($\zeta$) $=0$ and Im($\zeta$) $\neq 0$. It is evident that these regions of dynamic stability parameters are discrete, and the discrete nature of these stable regions arises directly from the resonant structure of the Bessel functions in $\rho_{2}$, which is a hallmark of Floquet engineering.

To verify the effectiveness of our perturbation analysis for correlated pair tunneling in the stability parameter regions far from resonance, we select the initial state $|0020\rangle$ and define the time-averaged probability of the unpaired state $|1010\rangle$, $\bar{P_7} = \frac{1}{\tau}\int_{0}^{\tau}P_{7}dt$. If the time averaged probability of the unpaired state in the stability parameter region is less than 0.02, we call this region "far from the resonance region", corresponding to stable correlated pair tunneling. In figure 1(a), when the parameters are taken in the stability region between two dashed lines near an integer, $\bar{P_7}$ is greater than 0.02. Otherwise, $\bar{P_7}$ is less than 0.02, indicating that it is far away from the resonance region. In figure 1(b), we selected $2f/\omega=1.6$ and $2f/\omega=5.57$ as the function curves of $\bar{P_7}$ with $2U_{1}/\omega$ around $2U_{1}/\omega=1$, respectively. From Figure 1(b), it can be seen that the peak of the curve corresponding to $2f/\omega=5.57$ is sharper and its near resonance range is narrower compared to $2f/\omega=1.6$. We take $2f/\omega=1.6$ for figure 1(c) and figure 1(d), and then take $2U_{1}/\omega=1.4$ for the far resonance region and $2U_{1}/\omega=1.1$ for the near resonance region, respectively, to represent the evolution of probabilities over time. The circle points label the analytical results from equation (20), and the solid curves denote the numerical correspondences obtained from equation (5). It is evident that in the far resonance region, only stable spin-conserving tunneling occurs between paired states, and the numerical and analytical solutions agree well, as shown in figure 1(c). In the near resonance region, the stable spin-conserving tunneling between paired and unpaired states occur simultaneously, and there is a deviation between the analytical and numerical solutions, as shown in figure 1(d). This proves the effectiveness of second-order perturbation analysis in the far resonance region.

\begin{figure}
  \includegraphics[width=0.48\linewidth]{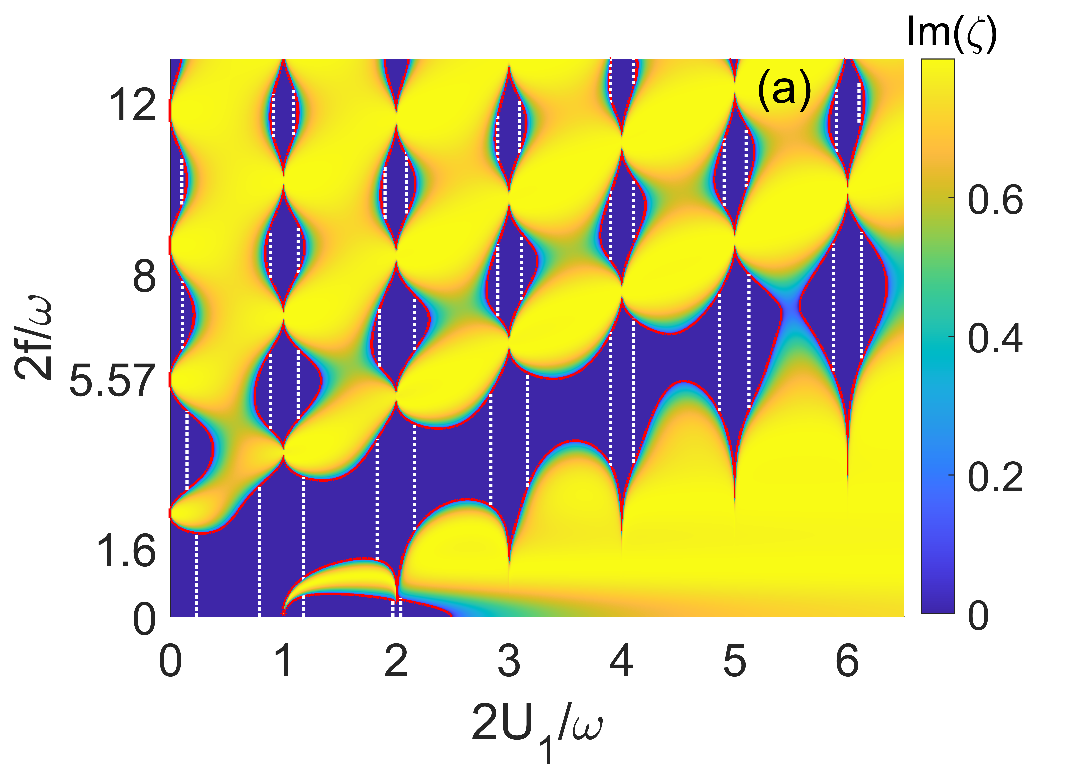}
  \includegraphics[width=0.48\linewidth]{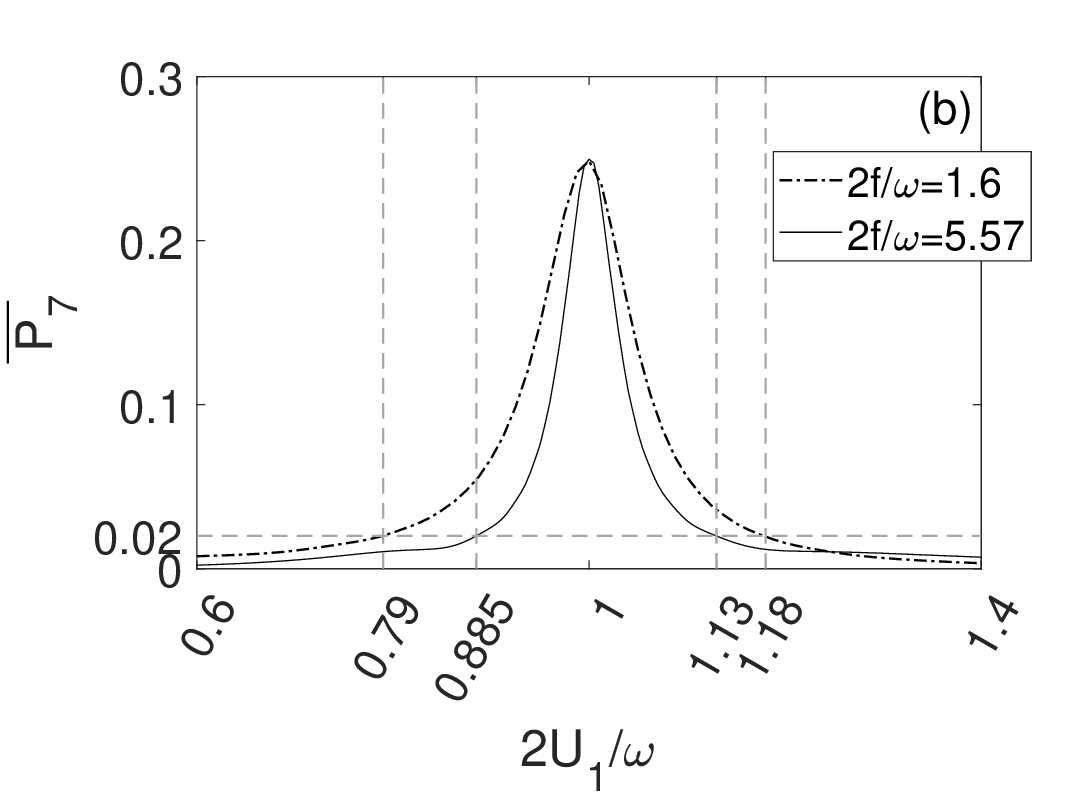}
  \includegraphics[width=0.48\linewidth]{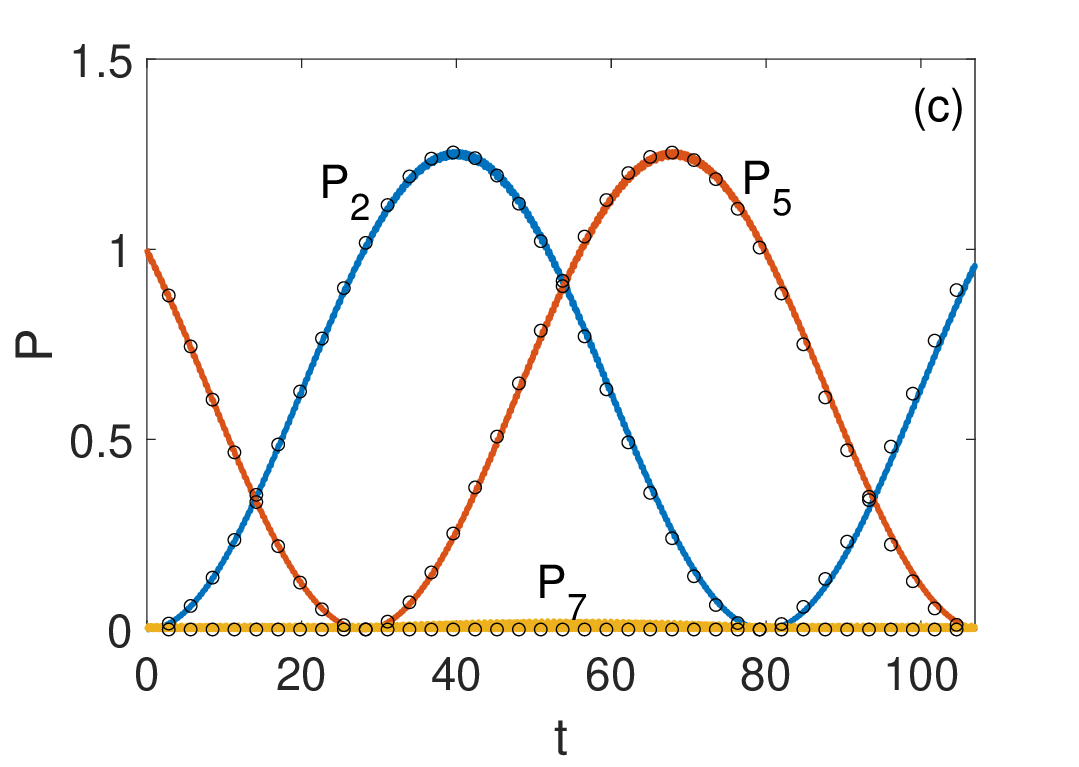}
  \includegraphics[width=0.48\linewidth]{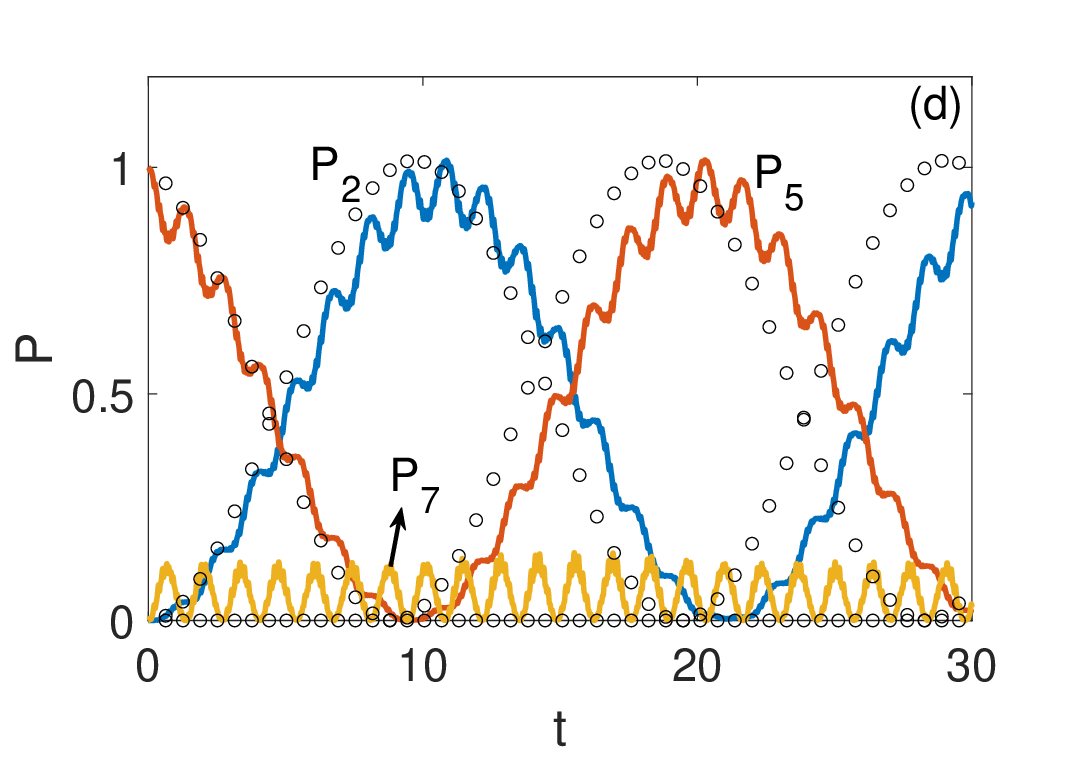}
  \caption{(a) Im($\zeta$) as a function of $2U_{1}/\omega$ and $2f/\omega$. (b) The time-averaged probabilities of the unpaired state $\bar{P_7}$ as a function of $2U_{1}/\omega$ for $2f/\omega=1.6$ (dash-dotted line) and $2f/\omega=5.57$ (solid line), respectively. (c)-(d) The time evolutions of the probabilities for (c) $2f/\omega=1.6$, $2U_{1}/\omega=1.4$, and (d) $2f/\omega=1.6$, $2U_{1}/\omega=1.1$. The initial state of the system is state $|0020\rangle$, and the other parameters are chosen as $\nu=\alpha=1$, $\delta=0$, $\omega=40$, $\Omega=40$, and $\beta_{1}=\beta_{2}=0.01$. Hereafter, circle points label the analytical results and solid curves denote the numerical correspondences. All parameters adopted in these figures are dimensionless.}
\end{figure}

\subsubsection{Stability under unbalanced gain and loss}

For unbalanced gain and loss, namely, $\beta_{1}<\beta_{2}$, stability of this system can be achieved when the parameters satisfy the balance condition
\begin{equation}
\begin{aligned}
\beta_{1}\beta_{2}=\frac{\nu^{4}\rho_{2}^{2}}{\omega^{2}}.
\end{aligned}
\end{equation}
Under this condition, the quasienergies in equation (21) become
\begin{equation}
\begin{aligned}
E_{1}&=-\frac{4\nu^{2}\rho_{1}}{\omega}+i(\beta_{1}-\beta_{2}),\\
E_{2}&=2U_{1}+\frac{2\nu^{2}\rho_{1}}{\omega}+2i(\beta_{1}-\beta_{2}),\\
E_{3}&=2U_{1}+\frac{2\nu^{2}\rho_{1}}{\omega}.
\end{aligned}
\end{equation}

\begin{figure}
  \includegraphics[width=0.48\linewidth]{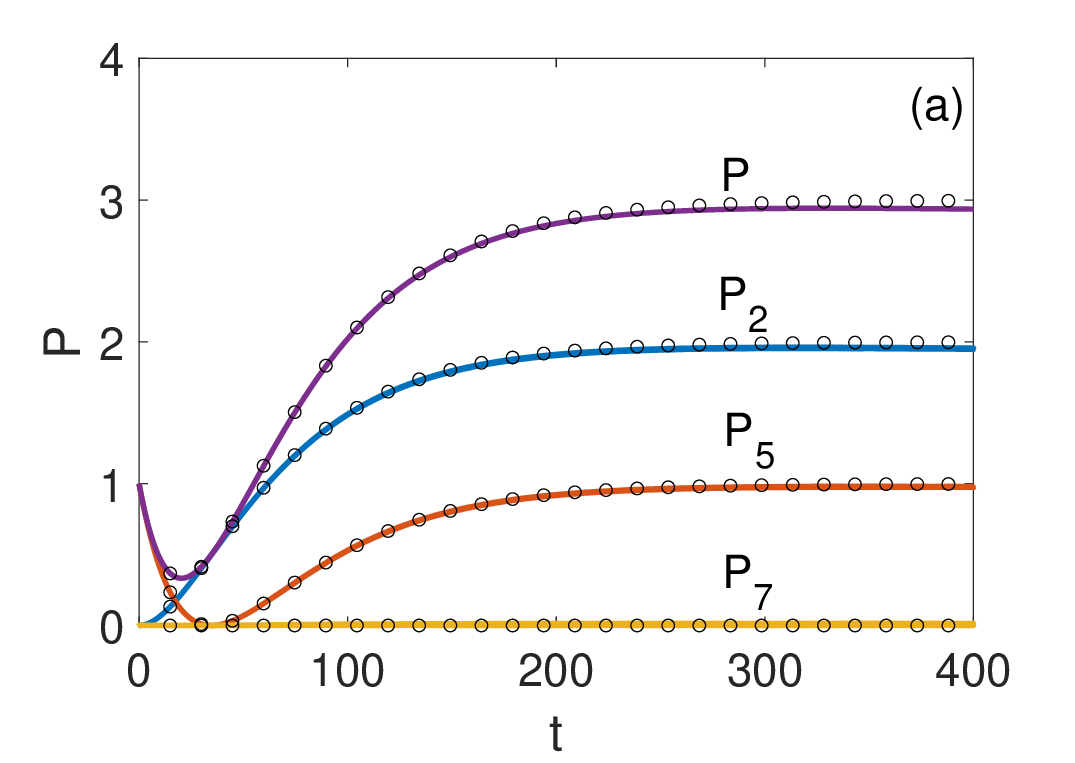}
  \includegraphics[width=0.48\linewidth]{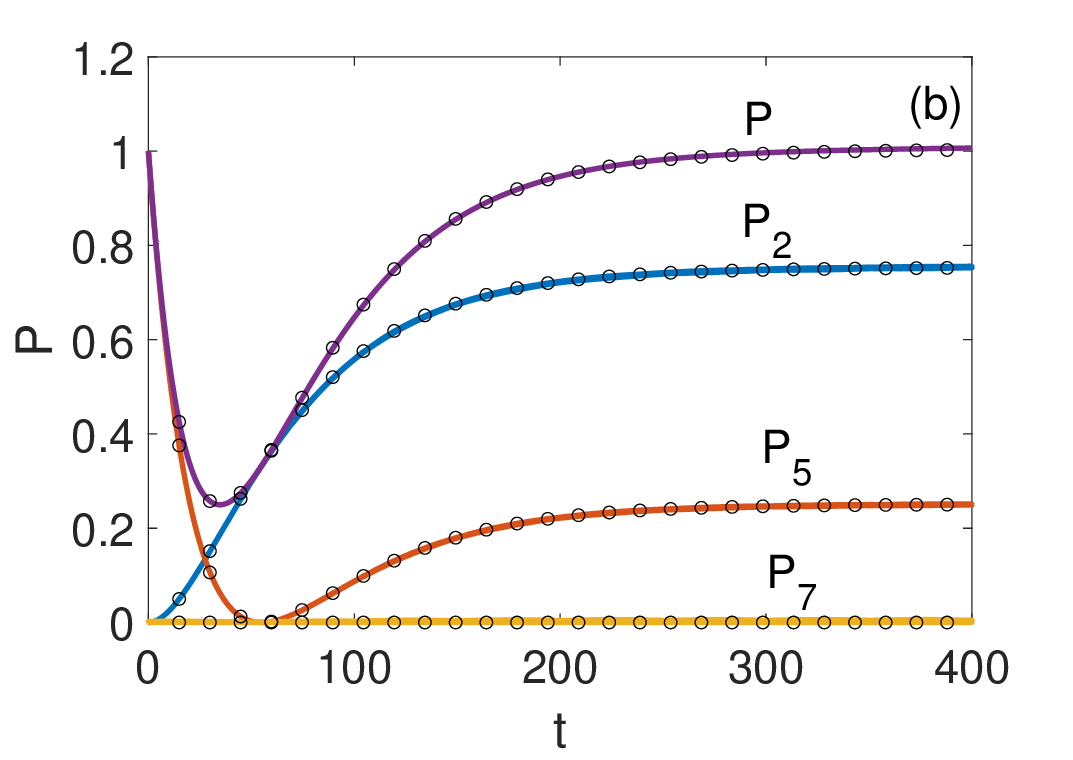}
  \caption{The time evolutions of the probabilities $P_k$ ($k=2,5,7$) and the total probability $P$ for different gain and loss coefficients. (a) $\beta_{1}=0.01$, $\beta_{2}=0.02$, $f=90.491$; (b) $\beta_{1}=0.005$, $\beta_{2}=0.015$, $f=78.28$. The initial state of the system is state $|0020\rangle$, and the other parameters are chosen as $\nu=\alpha=1$, $\delta=0$, $\omega=\Omega=40$, and $U_{1}=70$.}
\end{figure}

Since $\beta_{1}<\beta_{2}$, the imaginary parts of $E_1$ and $E_2$ are negative, while $E_3$ is real, satisfying the stability criterion\cite{Luo2020}. We set the initial state of the system $|0020\rangle$ and take the parameters that satisfy equation (24) to plot the evolutions of the probabilities $P_k$ and the total probability $P=\sum_{k} P_{k}$ ($k=2,5,7$) over time, as shown in figure 2.
Figures 2(a) and 2(b) show that the probabilities $P_{k}$ and the total probability $P$ tend to constant values at long times, confirming the system's stability and the validity of our analytical predictions. Specifically, when $\beta_{2}/\beta_{1}=3$, the final total probability is equal to the initial probability, and both are equal to one, which is consistent with that in existing literatures\cite{Luo2020, xiao85}.

\subsection{Interwell spin-flipping tunneling}

For half-integer $\alpha$ (e.g., $\alpha=0.5$) and $\delta=0$, the Hamiltonian of the system (1) reduces to
\begin{equation}
\begin{aligned}
\hat{H}&=-\nu(\hat{a}^{\dagger}_{1\downarrow}\hat{a}_{2\uparrow}-\hat{a}^{\dagger}_{1\uparrow}\hat{a}_{2\downarrow}+H.c.)\\
&+\sum_{j=1,2}[\Omega (t)\hat{a}^{\dagger}_{j}\hat{\sigma}_{z}\hat{a}_{j}+f_{j}(t)\hat{a}^{\dagger}_{j}\hat{a}_{j}]\\
&+\sum_{j=1,2}\sum_{\sigma\sigma^{\prime}}U_{\sigma\sigma^{\prime}}\hat{a}^{\dagger}_{j\sigma}\hat{a}^{\dagger}_{j\sigma^{\prime}}\hat{a}_{j\sigma^{\prime}}\hat{a}_{j\sigma}.
\end{aligned}
\end{equation}
So that the dominant tunneling process is the spin-flipping tunneling. Starting from the initial state $|0020\rangle$, the dynamics is confined to the subspace $\{|0020\rangle,|0110\rangle,|0200\rangle\}$. Following the same multiple-scale analysis as in Sec.II.A, we derive the effective second-order equations for this channel
\begin{equation}
\begin{aligned}
i\frac{dA_{5}}{dt}&=\frac{2\nu^{2}}{\omega}(A_{3}\rho_{4}+A_{5}\rho_{3})-2i\beta_{2}A_{5},\\
i\frac{dA_{9}}{dt}&=-\frac{4\nu^{2}}{\omega}A_{9}\rho_{3}+(i\beta_{1}-i\beta_{2})A_{9},\\
i\frac{dA_{3}}{dt}&=\frac{2\nu^{2}}{\omega}(A_{3}\rho_{3}+A_{5}\rho_{4})+2i\beta_{1}A_{3}.
\end{aligned}
\end{equation}
Here, the effective couplings are modified by the ac driving and Zeeman field
\begin{equation}
\begin{aligned}
\rho_{3}&=\sum_{p}\frac{J_{p}^{2}(\frac{2f}{\omega}-\frac{2\Omega}{\omega})}{p+\frac{2U_{1}}{\omega}},\\
\rho_{4}&=\sum_{p}\frac{J_{p}(\frac{2f}{\omega}-\frac{2\Omega}{\omega})J_{-p}(\frac{2f}{\omega}-\frac{2\Omega}{\omega})}{p+\frac{2U_{1}}{\omega}}.
\end{aligned}
\end{equation}
The corresponding quasienergies are
\begin{equation}
\begin{aligned}
E'_{1}&=i(\beta_{1}-\beta_{2})-\frac{4\nu^{2}\rho_{3}}{\omega},\\
E'_{2}&=2U_{1}+i(\beta_{1}-\beta_{2})+\frac{2\nu^{2}\rho_{3}}{\omega}-\xi,\\
E'_{3}&=2U_{1}+i(\beta_{1}-\beta_{2})+\frac{2\nu^{2}\rho_{3}}{\omega}+\xi,
\end{aligned}
\end{equation}
with $\xi=\sqrt{(\frac{2\nu^{2}\rho_{4}}{\omega})^{2}-(\beta_{1}+\beta_{2})^{2}}$.

\subsubsection{Stability under balanced gain and loss}

\begin{figure}
  \includegraphics[width=0.48\linewidth]{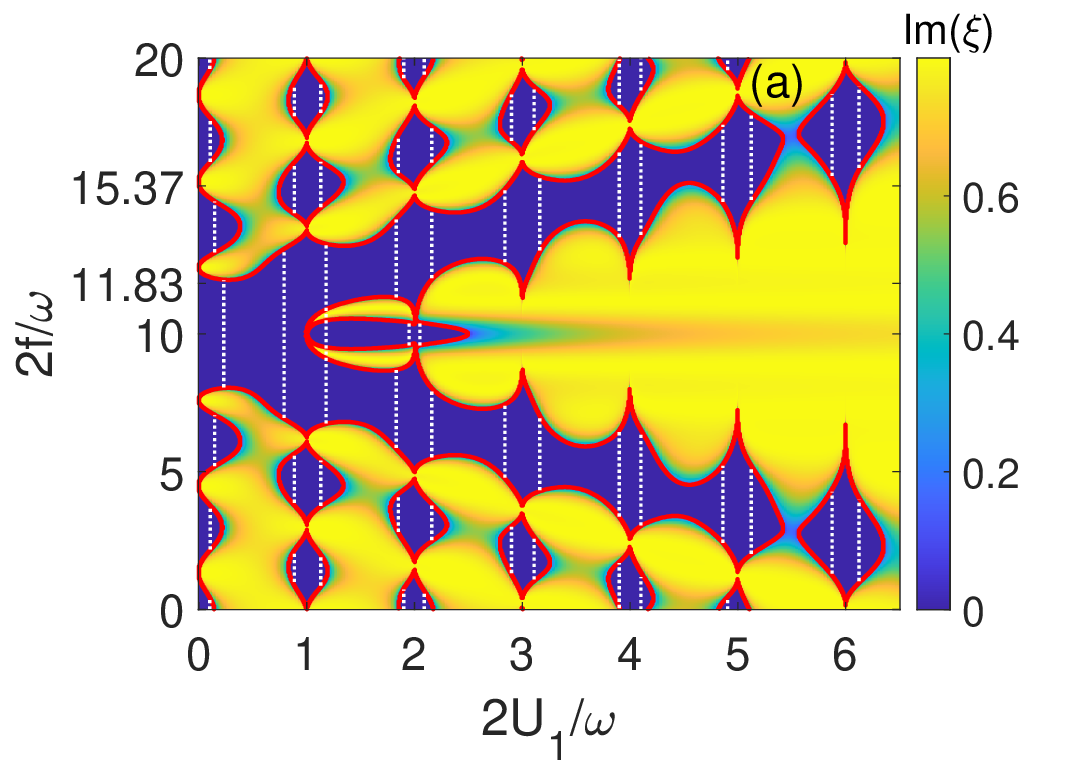}
  \includegraphics[width=0.48\linewidth]{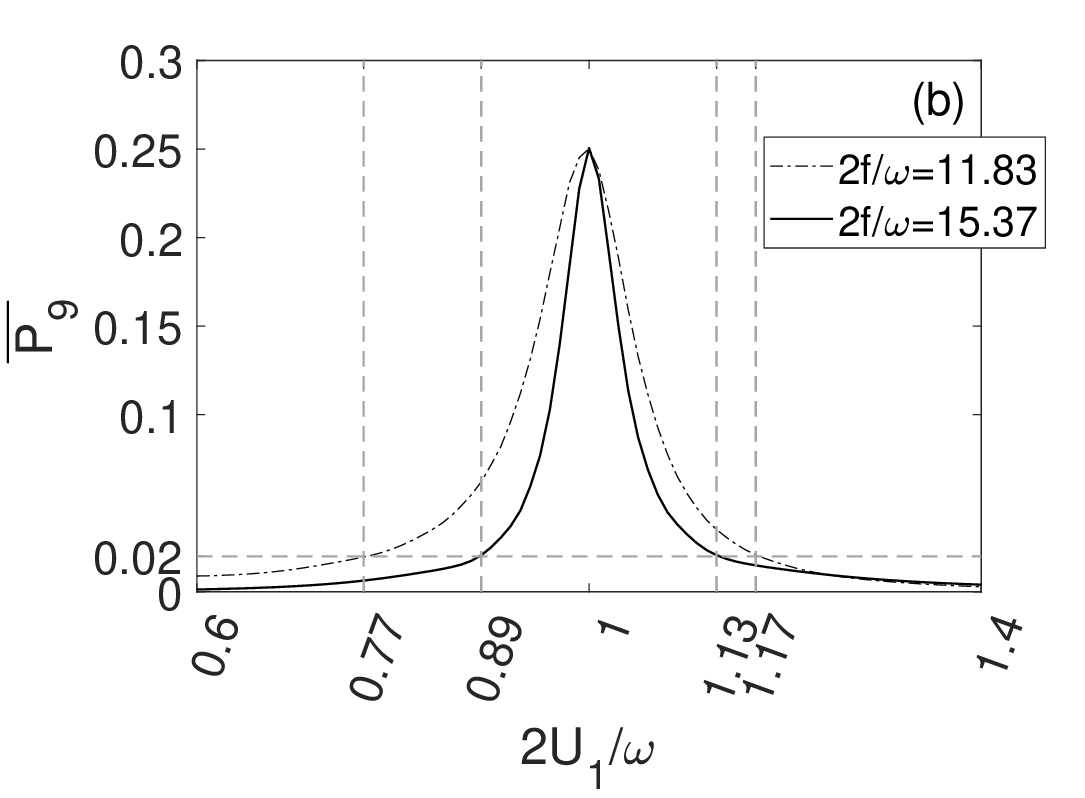}
  \includegraphics[width=0.48\linewidth]{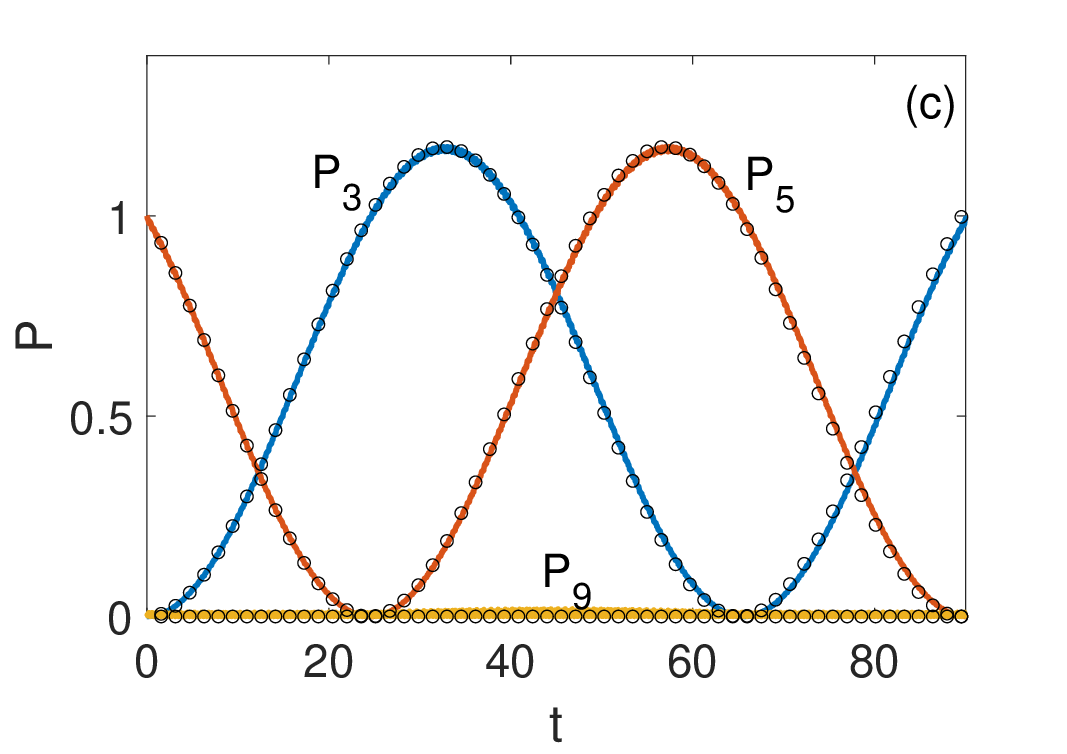}
  \includegraphics[width=0.48\linewidth]{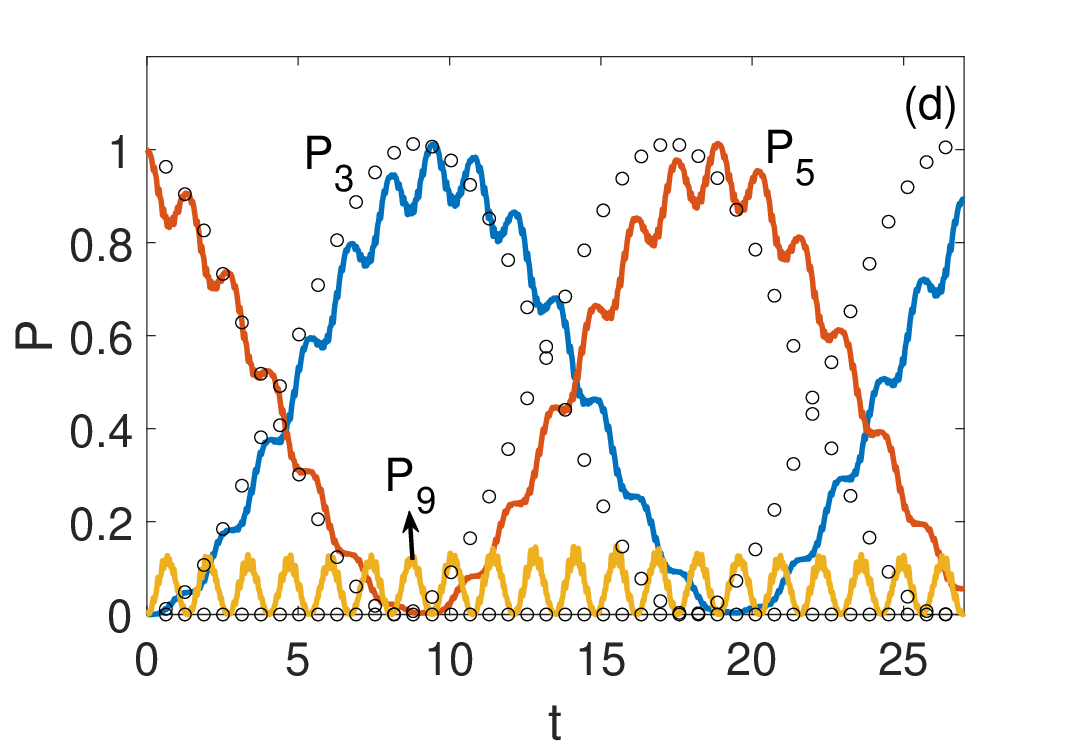}
  \caption{(a) Im($\xi$) as a function of $2U_{1}/\omega$ and $2f/\omega$. (b)The time-averaged probabilities of the unpaired state $\bar{P_{9}}$ as a function of $2U_{1}/\omega$ for $2f/\omega=11.83$ (dash-dotted line) and $2f/\omega=15.37$ (solid line), respectively. (c)-(d) The time evolutions of probabilities for (c) $2f/\omega=11.83$, $2U_{1}/\omega=1.4$, and (d) $2f/\omega=11.83$, $2U_{1}/\omega=1.1$. The initial state of the system is state $|0020\rangle$, and the other parameters are chosen as $\nu=1$, $\alpha=0.5$, $\delta=0$, $\omega=40$, $\Omega=200$, and $\beta_{1}=\beta_{2}=0.01$.}
\end{figure}

For balanced gain and loss ($\beta_{1}=\beta_{2}=\beta$), the quasienergies in equation (29) become
\begin{equation}
\begin{aligned}
E'_{1}&=-\frac{4\nu^{2}\rho_{3}}{\omega},\\
E'_{2}&=2U_{1}+\frac{2\nu^{2}\rho_{3}}{\omega}-\xi,\\
E'_{3}&=2U_{1}+\frac{2\nu^{2}\rho_{3}}{\omega}+\xi,
\end{aligned}
\end{equation}
with $\xi=\sqrt{(\frac{2\nu^{2}\rho_{4}}{\omega})^{2}-(2\beta)^{2}}$. Stability again requires Im($\xi$)=0. In figure 3(a), we take the same parameters as figure 1(a) except for $\Omega=200$ and plot Im($\xi$) as a function of $2U_{1}/\omega$ and $2f/\omega$, where the red line represents the boundary between Im($\xi$) $=0$ and Im($\xi$) $\neq 0$. From figure 3(a), it can be seen that for spin-flipping tunneling, the stability parameter regions of the system are also discrete, but for $2f/\omega=2\Omega/\omega=10$ symmetry, which is absent in the spin-conserving case in figure 1(a). To understand the origin of this symmetry, we examine the expression for $\rho_{4}$ given in equation (28). For any point $(2U_1/\omega,2f/\omega)$ within the stable parameter region, its symmetric counterpart with respect to the axis $2f/\omega=2\Omega/\omega$ is $(2U_1/\omega,4\Omega/\omega-2f/\omega)$. Substituting this symmetric point into $\rho_{4}$ yields
\begin{align}
\rho_{4}^{\prime}&=\sum_{p}\frac{J_{p}(\frac{4\Omega}{\omega}-\frac{2f}{\omega}-\frac{2\Omega}{\omega})J_{-p}(\frac{4\Omega}{\omega}-\frac{2f}{\omega}-\frac{2\Omega}{\omega})}{p+\frac{2U_1}{\omega}}\nonumber\\
&=\sum_{p}\frac{J_{p}(\frac{2\Omega}{\omega}-\frac{2f}{\omega})J_{-p}(\frac{2\Omega}{\omega}-\frac{2f}{\omega})}{p+\frac{2U_1}{\omega}}.
\end{align}
Using the Bessel function property $J_{p}(-x)=(-1)^{p}J_{p}(x)$, we obtain
\begin{align}
\rho_{4}^{\prime}&=\sum_{p}\frac{(-1)^{p}J_{p}(\frac{2f}{\omega}-\frac{2\Omega}{\omega})(-1)^{-p}J_{-p}(\frac{2f}{\omega}-\frac{2\Omega}{\omega})}{p+\frac{2U_1}{\omega}}\nonumber\\
&=\sum_{p}\frac{J_{p}(\frac{2f}{\omega}-\frac{2\Omega}{\omega})J_{-p}(\frac{2f}{\omega}-\frac{2\Omega}{\omega})}{p+\frac{2U_1}{\omega}}=\rho_{4}.
\end{align}
Thus, $\rho_{4}^{\prime}=\rho_{4}$, which proves that the stability landscape is symmetric about $2f/\omega=2\Omega/\omega$. This symmetry arises from the even/odd properties of Bessel functions under argument sign change and is a direct consequence of the Floquet engineering of the driven system.
Here, we set the initial state $|0020\rangle$ and define the time-averaged probability of the unpaired state $|0110\rangle$, $\bar{P_9} = \frac{1}{\tau}\int_{0}^{\tau}P_{9}dt$. In figure 3(a), when the parameters are taken in the stable region between two dashed lines near the integer, $\bar{P_9}$ is greater than 0.02. Otherwise, $\bar{P_9}$ is less than 0.02, showing that it is far away from the resonance region and stable spin-flipping tunneling of correlated pair states occurs. In figure 3(b), we selected $2f/\omega=11.83$ and $2f/\omega=15.37$ as the function curves of $\bar{P_9}$ with $2U_{1}/\omega$ around $2U_{1}/\omega=1$, respectively. Figure 3(b) shows that the peak of the curve corresponding to $2f/\omega=15.37$ is sharper and its near resonance range is narrower compared to $2f/\omega=11.83$. In figures 3(c) and 3(d), we take $2f/\omega=11.83$, $2U_{1}/\omega=1.4$ (the far resonance region), and $2U_{1}/\omega=1.1$ (the near resonance region), respectively, to plot the evolution of probabilities over time. The circle points label the analytical results from equation (27), and the solid curves denote the numerical correspondences from equation (26). It is evident that the stable spin-flipping tunneling of correlated pair states occurs far from the resonance region, as shown in figure 3(c). While in the near resonance region, quantum transitions coexist between states $|0020\rangle$, $|0110\rangle$, and $|0200\rangle$, and the numerical results deviate from the analytical results, as shown in figure 3(d). This once again proves the effectiveness of second-order perturbation analysis for correlated pair tunneling in regions far from resonance.

\subsubsection{Stability under unbalanced gain and loss}

\begin{figure}
  \includegraphics[width=0.48\linewidth]{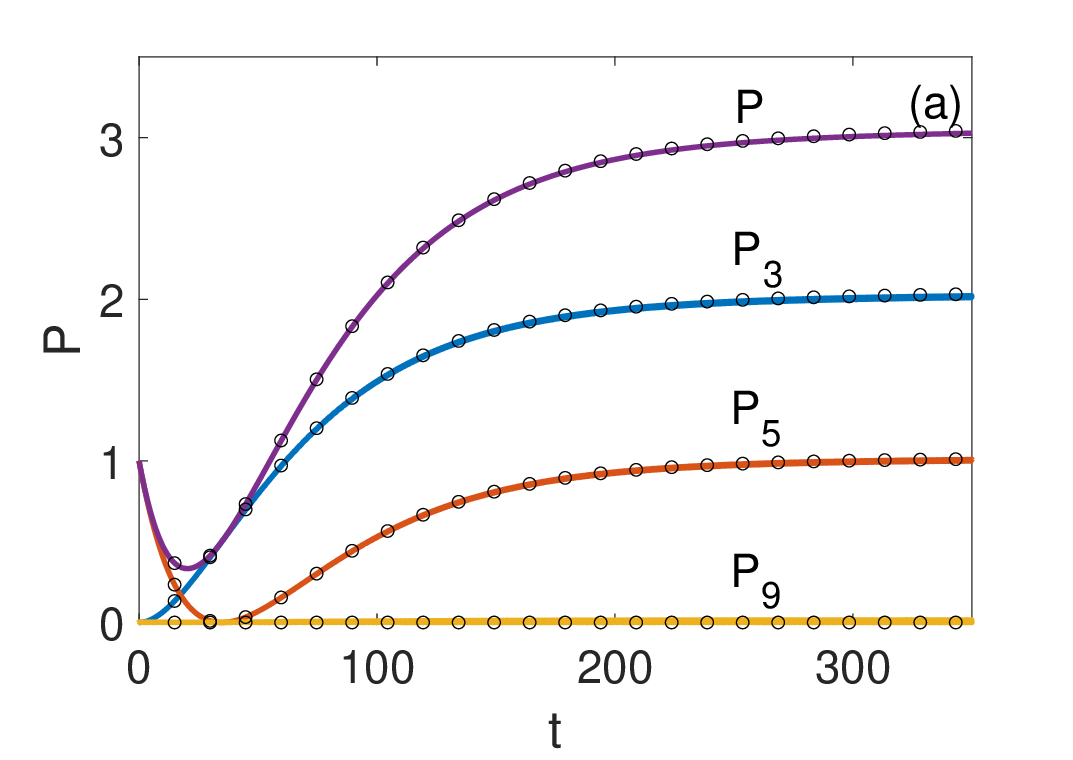}
  \includegraphics[width=0.48\linewidth]{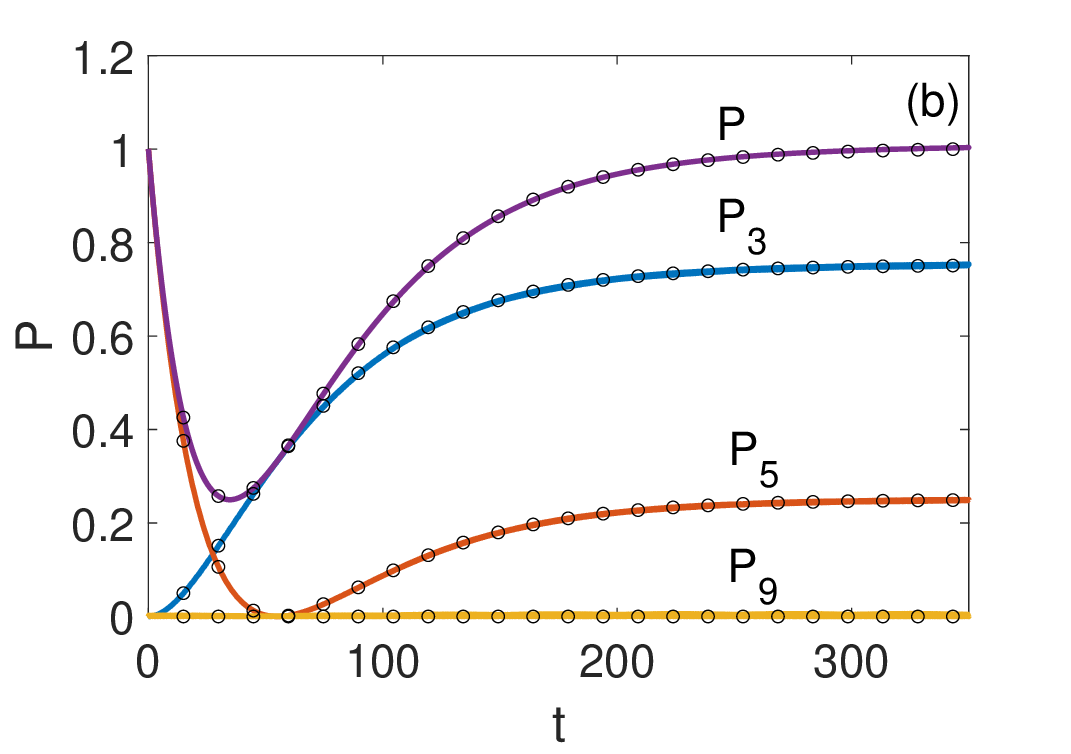}
  \caption{The time evolutions of the probabilities $P_{k}$ ($k=3,5,9$) and the total probability $P$ for different gain and loss coefficients. (a) $\beta_{1}=0.01$, $\beta_{2}=0.02$, $f=143.3$; (b) $\beta_{1}=0.005$, $\beta_{2}=0.015$, $f=118.28$. The initial state of the system is state $|0020\rangle$, and the other parameters are chosen as $\nu=1$, $\alpha=0.5$, $\delta=0$, $\omega=\Omega=40$, and $U_{1}=70$.}
\end{figure}

For unbalanced gain and loss ($\beta_{1}<\beta_{2}$), stability in the spin-flipping channel is achieved when the parameters satisfy the balance condition
\begin{equation}
\begin{aligned}
\beta_{1}\beta_{2}=\frac{\nu^{4}\rho_{4}^{2}}{\omega^{2}}.
\end{aligned}
\end{equation}
Under this condition, the quasienergies in equation (29) reduce to
\begin{equation}
\begin{aligned}
E'_{1}&=-\frac{4\nu^{2}\rho_{3}}{\omega}+i(\beta_{1}-\beta_{2}),\\
E'_{2}&=2U_{1}+\frac{2\nu^{2}\rho_{3}}{\omega}+2i(\beta_{1}-\beta_{2}),\\
E'_{3}&=2U_{1}+\frac{2\nu^{2}\rho_{3}}{\omega}.
\end{aligned}
\end{equation}
Obviously, the quasienergies meet stability condition (ii), and the system is stable. In figures 4(a) and 4(b), we set the initial state of the system to state $|0020\rangle$ and take two sets of parameters that satisfy equation (31) to plot the time evolutions of the probabilities $P_k$ and the total probability $P=\sum_{k} P_{k}$ ($k=3,5,9$). It can be seen that all the probabilities tend to remain constant after a period of evolution. Specially, in figure 4(b), when $\beta_2/\beta_1=3$, the final total probability is equal to the initial probability of one, which is consistent with the conclusion obtained from interwell spin-conserving tunneling.

\subsection{Intrawell spin-flipping}

In the subsection, we focus on the intrawell spin-flipping process of paired states. From equation (20) and equation (27), we find when $\alpha=1$ and $\rho_{2}=0$ or $\alpha=0.5$ and $\rho_{4}=0$, the interwell spin-conserving tunneling or interwell spin-flipping tunneling is forbidden, and the system exhibits only intrawell spin-flipping dynamics induced by the Raman coupling $\delta$. Due to the gain of well 1 and the dissipation of well 2,
when two bosons are initially completely trapped in well 1 or well 2, the probability of dynamic quantum transitions in the system will exponentially increase or decrease. Therefore, in order to achieve stable intrawell spin-flipping transitions of correlated pair states, the initial state of the system can only be a superposition state of two bosons in well 1 and well 2. Here, we select a superposition state of $|0020\rangle$ and $|2000\rangle$ as an initial state and take the parameters $\alpha=1$ and $\rho_{2}=0$ as an example to study the stability intrawell spin-flipping of correlated pair states. The quantum dynamics of the system is confined to the subspace $\{|0020\rangle,|0011\rangle,|0002\rangle,|2000\rangle,|1100\rangle,|0200\rangle\}$. Following the same procedure, we obtain the second-order coupled equations as follow
\begin{equation}
\begin{aligned}
i\frac{dA_{1}}{dt}&=-\frac{\delta^{2}}{\omega}A_{1}\rho_{5}+2i\beta_{1}A_{1},\\
i\frac{dA_{2}}{dt}&=\frac{\delta^{2}}{2\omega}(A_{2}\rho_{5}+A_{3}\rho_{6})+2i\beta_{1}A_{2},\\
i\frac{dA_{3}}{dt}&=\frac{\delta^{2}}{2\omega}(A_{2}\rho_{6}+A_{3}\rho_{5})+2i\beta_{1}A_{3},\\
i\frac{dA_{4}}{dt}&=-\frac{\delta^{2}}{\omega}A_{4}\rho_{5}-2i\beta_{2}A_{4},\\
i\frac{dA_{5}}{dt}&=\frac{\delta^{2}}{2\omega}(A_{5}\rho_{5}+A_{6}\rho_{6})-2i\beta_{2}A_{5},\\
i\frac{dA_{6}}{dt}&=\frac{\delta^{2}}{2\omega}(A_{5}\rho_{6}+A_{6}\rho_{5})-2i\beta_{2}A_{6},
\end{aligned}
\end{equation}
with
\begin{equation}
\begin{aligned}
\rho_{5}=\sum_{p}\frac{J_{p}^{2}(\frac{2\Omega}{\omega})}{p+\frac{2U_{1}}{\omega}-\frac{2U_{2}}{\omega}},\\
\rho_{6}=\sum_{p}\frac{J_{p}(\frac{2\Omega}{\omega})J_{-p}(\frac{2\Omega}{\omega})}{p+\frac{2U_{1}}{\omega}-\frac{2U_{2}}{\omega}}.
\end{aligned}
\end{equation}

The corresponding Floquet quasienergies are
\begin{equation}
\begin{aligned}
E''_{1}&=2U_{2}+2i\beta_{1}-\frac{\delta^{2}}{\omega}\rho_{5},\\
E''_{2}&=2U_{1}+2i\beta_{1}+\frac{\delta^{2}}{2\omega}(\rho_{5}-\rho_{6}),\\
E''_{3}&=2U_{1}+2i\beta_{1}+\frac{\delta^{2}}{2\omega}(\rho_{5}+\rho_{6}),\\
E''_{4}&=2U_{2}-2i\beta_{2}-\frac{\delta^{2}}{\omega}\rho_{5},\\
E''_{5}&=2U_{1}-2i\beta_{2}+\frac{\delta^{2}}{2\omega}(\rho_{5}-\rho_{6}),\\
E''_{6}&=2U_{1}-2i\beta_{2}+\frac{\delta^{2}}{2\omega}(\rho_{5}+\rho_{6}).
\end{aligned}
\end{equation}

\begin{figure}
  \includegraphics[width=0.8\linewidth]{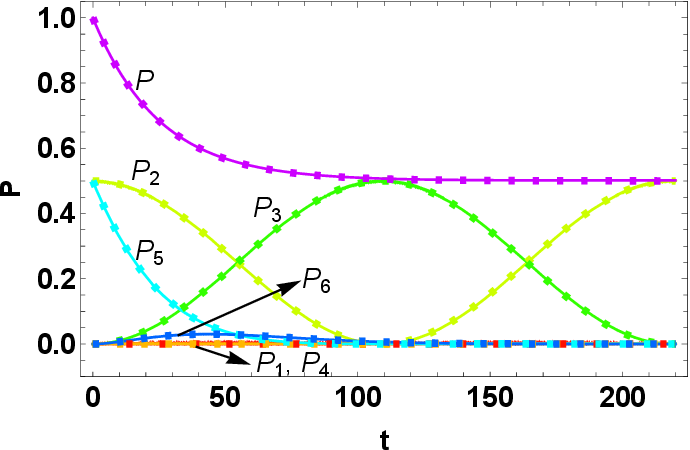}
  \caption{The time evolutions of the probabilities $P_{k}$ and the total probability $P=\sum_{k} P_{k}$ ($k=1,2,...,6$). The initial state of the system is state $|\psi(0)\rangle=\frac{1}{\sqrt{2}}|0020\rangle+\frac{1}{\sqrt{2}}|2000\rangle$. The parameters are chosen as $\nu=\alpha=\delta=1$, $\omega=40$, $U_{1}=51.4586$, $U_{2}=24$, $f=110$, $\Omega=40$, $\beta_{1}=0$, and $\beta_{2}=0.1$.}
\end{figure}

As derived in the full six-mode analysis, this allows for a scenario where, by setting $\beta_{1}=0$, three of the quasienergies become real and the imaginary parts of the other quasienergies are less than zero, enabling stable dynamics. In figure 5, we set the initial state $|\psi(0)\rangle=\frac{1}{\sqrt{2}}|0020\rangle+\frac{1}{\sqrt{2}}|2000\rangle$ and take the parameters that satisfy $\rho_{2}=0$ to plot the time evolutions of the probabilities $P_{k}$ and the total probability $P=\sum_{k} P_{k}$ ($k=1,2,...,6$). It can be seen that the intrawell spin-flipping of correlated pair states is stable. This indicates that in non-Hermitian systems, initial-state coherence can be harnessed to selectively populate subspaces that are immune to dissipation, thereby stabilizing dynamics that would otherwise be forbidden.

\section{Conclusions}
In summary, we have developed an analytical framework for understanding and stabilizing correlated pair tunneling of two spin-orbit-coupled bosons in a driven non-Hermitian double well. Our work yields three principal findings of broad significance. First, we demonstrate that the stability of correlated tunneling is governed by a competition between coherent Floquet-renormalized coupling and local dissipation, leading to well-defined, engineerable stability islands in parameter space. Second, we uncover a hidden symmetry in the spin-flipping channel, a direct consequence of the interplay between driving and SOC, which provides an additional knob for coherent control. Third, we establish a fundamental link between initial-state coherence and dynamical stability, demonstrating that a coherent superposition can shield a subsystem from dissipation, thereby enabling dynamics that are forbidden for a simple Fock state. We expect that these finds will expand the possibilities for manipulating the spin dynamics of bound states in non-Hermitian many-body quantum systems.

\section*{ACKNOWLEDGMENTS}

This work was supported by the National Natural Science Foundation of China under Grant No. 11747034, the Scientific Research Foundation of Ningxia Education Department under Grant No. NYG2024202, National Students' Platform for Innovation and Entrepreneurship Training Program under Grant No. 202510542020, and "Tenglong" Innovative Talent Fund of Hunan Normal University under Grant No. 2025TL104.

\end{document}